\documentclass[12pt]{article}

\usepackage{scicite}
\usepackage{times}
\usepackage{graphicx}
\usepackage{xcolor}
\usepackage{upgreek}

\newenvironment{sciabstract}{%
\begin{quote} \bf}
{\end{quote}}

\title{A massive stellar bulge in a regularly rotating galaxy 1.2 billion years after the Big Bang}

\author
{Federico Lelli,$^{1, 2, \ast}$, Enrico M. Di Teodoro$^{3}$, Filippo Fraternali$^{4}$, Allison W.S. Man$^{5}$,\\
Zhi-Yu Zhang$^{6}$, Carlos De Breuck$^{7}$, Timothy A. Davis$^{1}$, Roberto Maiolino$^{8, 9}$\\
\normalsize{$^{1}$School of Physics and Astronomy, Cardiff University, Cardiff CF24 3AA, UK}\\
\normalsize{$^{2}$Arcetri Astrophysical Observatory, INAF, Florence 50125, Italy; $^\ast$e-mail: federico.lelli@inaf.it}\\
\normalsize{$^{3}$Department of Physics \& Astronomy, Johns Hopkins University, Baltimore MD 21218, USA}\\
\normalsize{$^{4}$Kapteyn Astronomical Institute, University of Groningen, Groningen 9700 AV, The Netherlands}\\
\normalsize{$^{5}$Dunlap Institute for Astronomy \& Astrophysics, University of Toronto, Toronto ON M5S 3H4, Canada}\\
\normalsize{$^{6}$School of Astronomy and Space Science, Nanjing University, Nanjing 210023, P.R. China}\\
\normalsize{$^{7}$European Southern Observatory, Germany Headquarters, Garching bei M\"unchen  85748, Germany}\\
\normalsize{$^{8}$Kavli Institute for Cosmology, University of Cambridge, Cambridge CB3 0HA, UK}\\
\normalsize{$^{9}$Cavendish Laboratory, University of Cambridge, Cambridge CB3 0HE, UK}\\
}
\date{}

\newcommand{\cii} {{\rm [C\,{\footnotesize\rm II}]}}

\begin{document} 


\baselineskip24pt


\maketitle 

\newpage

\begin{sciabstract}
Cosmological models predict that galaxies forming in the early Universe experience a chaotic phase of gas accretion and star formation, followed by gas ejection due to feedback processes. Galaxy bulges may assemble later via mergers or internal evolution. Here we present submillimeter observations (with spatial resolution of 700 parsecs) of ALESS\,073.1, a starburst galaxy at\,$z\simeq5$\,when the Universe was 1.2 billion years old. This galaxy's cold gas forms a regularly rotating disk with negligible noncircular motions. The galaxy rotation curve requires the presence of a central bulge in addition to a star-forming disk. We conclude that massive bulges and regularly rotating disks can form more rapidly in the early Universe than predicted by models of galaxy formation.
\end{sciabstract}

In the standard $\Lambda$ cold dark matter ($\Lambda$CDM) cosmological model, galaxies form inside dark matter (DM) halos when primordial gas cools, collapses, and begins star formation \cite{White1978}. At early times, when the Universe was only 10$\%$ of its current age (redshift $z\simeq5$), $\Lambda$CDM models predict that galaxies undergo a turbulent phase of gas accretion from the cosmic web, leading to violent bursts of star formation and black hole growth \cite{Dekel2006}, rapidly followed by massive gas outflows due to feedback from supernovae and active galactic nuclei (AGN). Strong stellar and AGN feedback are needed in $\Lambda$CDM models to reproduce the observed number of galaxies per unit volume per unit mass (the galaxy stellar mass function) and to quench star formation in the most massive DM halos, reproducing the observed population of quiescent early-type galaxies at $z=0$ \cite{Vogelsberger2014}. In the hierarchical $\Lambda$CDM scenario, galaxy bulges (the spheroidal stellar components in the central parts of massive galaxies) are expected to form either after the merger of two galaxies of similar mass, or via secular dynamical processes within the stellar disk \cite{Kormendy2004}. Observation of galaxies at high redshifts enables us to test these scenarios by searching for these physical processes in action.

The 158-$\upmu$m emission line of singly ionized carbon, \cii, is a major coolant of the interstellar medium of galaxies and traces a combination of atomic and molecular hydrogen phases. The \cii\ line has been observed up to $z\simeq7.5$ \cite{Banados2019} and several rotating disks have been identified at $z>3$ \cite{DeBreuck2014, Jones2017, Pensabene2020}. Most existing observations only marginally resolve the \cii\ distribution and kinematics, and thus cannot distinguish between rotating disks, galaxy mergers, or gas inflows and outflows. Limited spatial resolution has prevented the measurement of well-sampled rotation curves (with more than six independent elements), which can be used to measure different mass components by fitting dynamical models.

We used the Atacama Large Millimeter/submillimeter Array (ALMA) to observe a star-forming galaxy at $z=4.75$ \cite{DeBreuck2014} at a spatial resolution of $\sim$0.11$''$, corresponding to 0.7 kpc in a $\Lambda$CDM cosmology \cite{Supp}. The target, ALESS\,073.1 (LESS J033229.3-275619), is a strong sub-millimeter source \cite{Coppin2009}. Its large far-infrared luminosity suggests an ongoing starburst with an estimated star-formation rate of $\sim$1000 solar masses (M$_\odot$) per year \cite{Coppin2009, Gilli2014}. ALESS\,073.1 contains a dust-obscured AGN \cite{Gilli2011}. These extreme properties indicate that the galaxy may drive a massive gas outflow \cite{Gilli2014}. Previous \cii\ observations with ALMA indicated a rotating disk \cite{DeBreuck2014}, but the spatial resolution (0.5$''$ corresponding to $\sim$3 kpc) was insufficient to determine its detailed properties or identify gas inflows and outflows.

\begin{figure*}
\includegraphics[width=16cm]{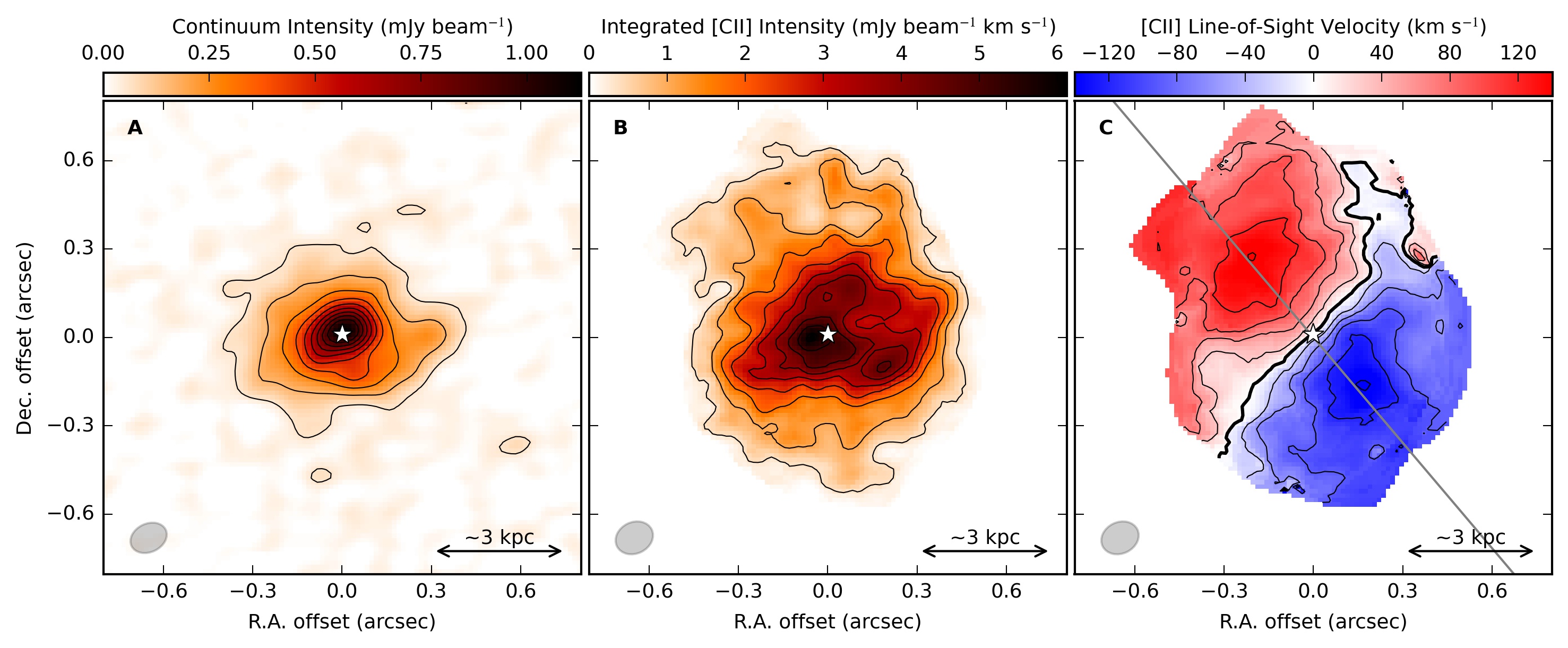}
\caption{\textbf{ALMA observations of ALESS\,073.1.} (\textbf{A}) Continuum emission at 160 $\upmu$m (rest-frame) tracing dust heated by young stars, (\textbf{B}) \cii\ intensity map tracing cold gas, and (\textbf{C}) \cii\ velocity field showing a rotating disk. North is up and east is left. The kinematic center, located at a Right Ascension (R.A.) of 03$^{\rm h}$ 32$^{\rm m}$ 29.295$^{\rm s}$ and Declination (Dec.) of $-$27$^\circ$ 56$'$ 19.60$''$, is represented by a white star. The beam size is plotted as the grey ellipse in the bottom-left corner. The physical scale is indicated by the scale bar in the bottom-right corner. In (A), iso-emission contours range from $0.055$ to 1 mJy\,beam$^{-1}$ (where 1 mJy = 10$^{-29}$ W m$^{-2}$ Hz$^{-1}$) in steps of 0.11 mJy\,beam$^{-1}$. In (B), iso-emission contours range from 0.35 to 6 mJy\,beam$^{-1}$ km\,s$^{-1}$ in steps of 0.7 mJy\,beam$^{-1}$ km\,s$^{-1}$. In (A) and (B), the lowest iso-emission contour corresponds to a signal-to-noise ratio of $\sim$3. In (C), iso-velocity contours range from $-120$ to $+120$ km\,s$^{-1}$ in steps of 30 km\,s$^{-1}$, the bold contour indicates the systemic velocity (set to zero), and the grey line shows the kinematic major axis.}
\end{figure*}

Figure 1A shows the continuum map at rest-frame 160 $\mu$m, which traces dust heated by the star-formation activity. Figure 1B shows the integrated \cii\ intensity map, which traces the cold gas distribution. The \cii\ emission is about twice as extended as the dust emission and shows an asymmetry to the northeast. Similar lopsided gas disks are common in the nearby Universe, forming nearly half of the local population of atomic gas disks \cite{Sancisi2008}.

The \cii\ velocity field (moment-one map; Fig. 1C) shows a regularly rotating disk. Rotating disks and galaxy mergers can be difficult to distinguish when observed at low spatial resolutions \cite{Sweet2019}. This is not the case for ALESS\,073.1 because the area with detectable \cii\ emission is covered by $\sim$45 resolution elements. The kinematic major axis is perpendicular to the kinematic minor axis, implying that noncircular motions are negligible in the inner parts of the gas disk \cite{Fraternali2001}. The gas kinematics are less regular in the northeastern extension: This may be due to recently accreted gas, a warped outer disk, or streaming motions along a spiral arm \cite{Supp}. The limited signal-to-noise ratio in the outer regions does not allow us to distinguish between these possibilities. Overall, the kinematic regularity of the \cii\ disk is unexpected for a starburst AGN-host galaxy at $z\simeq0$.

\begin{figure*}
\includegraphics[width=16cm]{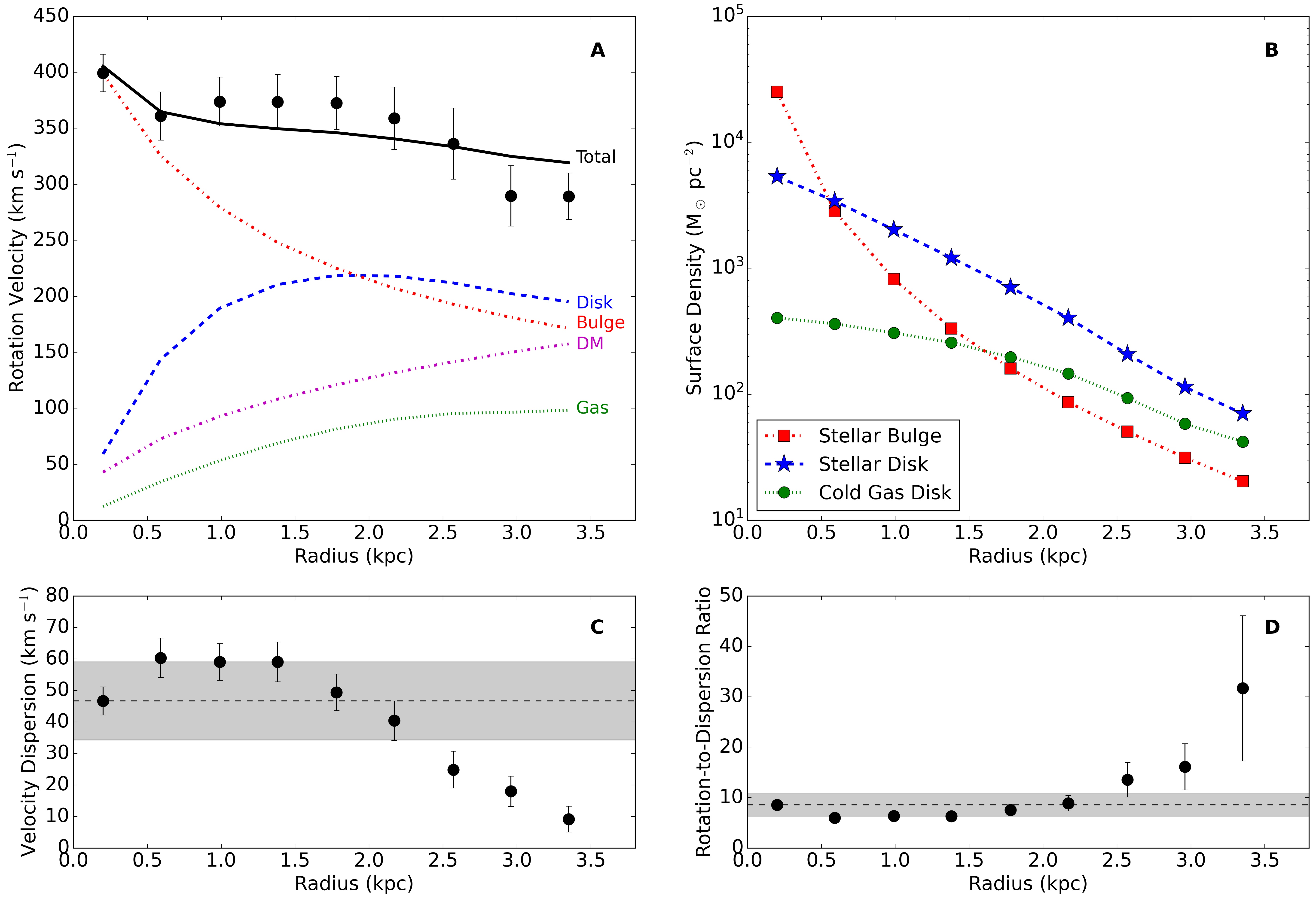}
\caption{\textbf{Mass model of ALESS\,073.1.}
(\textbf{A}): The observed rotation curve (black circles) adopting the best-fitting inclination of 22$^{\circ}$. The model rotation curve (black solid line) is the total of contributions from the stellar bulge (red dot-dashed line), stellar disk (blue dashed line), cold gas disk (green dotted line), and DM halo (magenta dash-dotted line). (\textbf{B}): Surface density profiles of the bulge (red squares), stellar disk (blue stars), and cold gas disk (green circles) adopting the best-fitting masses \cite{Supp}. The bulge profile assumes a de Vaucouleurs' distribution \cite{DeVauc1948}. The stellar and gas disk profiles are obtained from azimuthal averages of the dust continuum and \cii\ intensity maps, respectively. The uncertainties on these profiles are dominated by the systematic uncertainty on the normalizing masses \cite{Supp}. (\textbf{C}): \cii\ velocity dispersion profile. (\textbf{D}): Disk rotational support versus radius. In (C) and (D), the dashed line and grey band show the median value and median absolute deviation, respectively. In (A), (C), and (D), the error bars correspond to $\pm$1$\sigma$ uncertainties \cite{Supp}.}
\end{figure*}

We model the \cii\ kinematics using the software \textsc{$^{\rm 3D}$Barolo}\cite{DiTeodoro2015}, which fits the observational data with a rotating disk model \cite{Supp}. This determines the \cii\ surface brightness profile, the rotation curve, and the intrinsic velocity dispersion profile. The best-fitting rotation curve (Fig. 2A) rises steeply in the central parts and declines across the disk. Similar rotation curves are observed in bulge-dominated disk galaxies in the nearby Universe \cite{Casertano1991, Noordermeer2007, Lelli2016}, so there might be a bulge component in ALESS\,073.1. The normalization of the rotation curve depends on the disk inclination, which we treat as a free parameter below.

We use the derived rotation curve to constrain the mass distribution within the galaxy. Our fiducial mass model has four components: cold gas disk, stellar disk, stellar bulge, and DM halo. We also investigate mass models with three components, which we find either cannot fit the observations or are less plausible in the $\Lambda$CDM paradigm \cite{Supp}. We assume that the \cii\ density profile traces the distribution of the cold gas disk, whereas the dust density profile traces the distribution of the stellar disk (Fig. 2B), because dust absorbs ultraviolet radiation from young stars and re-emits it at far-infrared wavelengths. The two additional components (bulge and DM halo) are modeled with analytic functions \cite{Supp}. The bulge component implicitly includes the contribution of the central super-massive black hole, which contributes $<$10$\%$ of the central mass in high-$z$ galaxies \cite{Pensabene2020}.

For the best-fitting inclination of 22$^{\circ}$, the \cii\ disk rotates at velocities ($V_{\rm rot}$) of 300 to 400 km\,s$^{-1}$ (Fig. 2A). Similar rotational speeds are observed in the most massive early-type galaxies at $z=0$ that host molecular gas disks \cite{Davis2016}, so we conclude that they are the likely descendants of galaxies similar to ALESS\,073.1. The intrinsic velocity dispersion ($\sigma_{V}$) reaches 50 to 60 km s$^{-1}$ at small radii and decreases to $\sim$10 km s$^{-1}$ in the outer parts (Fig. 2C). The ratio of rotation velocity to velocity dispersion (Fig. 2D) is $\sim$10 within 2.5 kpc, similar to that of cold gas disks at $z=0$ \cite{Wisnioski2015}. The \cii\ disk of ALESS\,073.1 is supported by rotation, not turbulence.

The bulge component of the model is necessary to reproduce the observed high rotation speeds at small radii. We obtain a bulge-to-total mass ratio $M_{\rm bul}/M_{\rm baryon} = 0.44$, where $M_{\rm baryon}$ is given by the sum of all baryonic (non-DM) components within 3.5 kpc. This ratio is nearly independent of the disk inclination because it is largely driven by the shapes of the observed rotation curve and of the baryonic gravitational contributions, not by their absolute normalizations. Baryons dominate the gravitational potential within 3.5 kpc, similar to massive galaxies at $z=0$ \cite{Lelli2016} and $z=1$ to 3 \cite{Lelli2018, Genzel2017}. In $\Lambda$CDM cosmology, DM halos are expected to have low concentrations at high $z$, whereas baryons can efficiently cool and collapse to the bottom of the potential well, reaching high concentrations. Thus, DM halos become gravitationally dominant at large radii.

\begin{figure*}
\includegraphics[width=15.5cm]{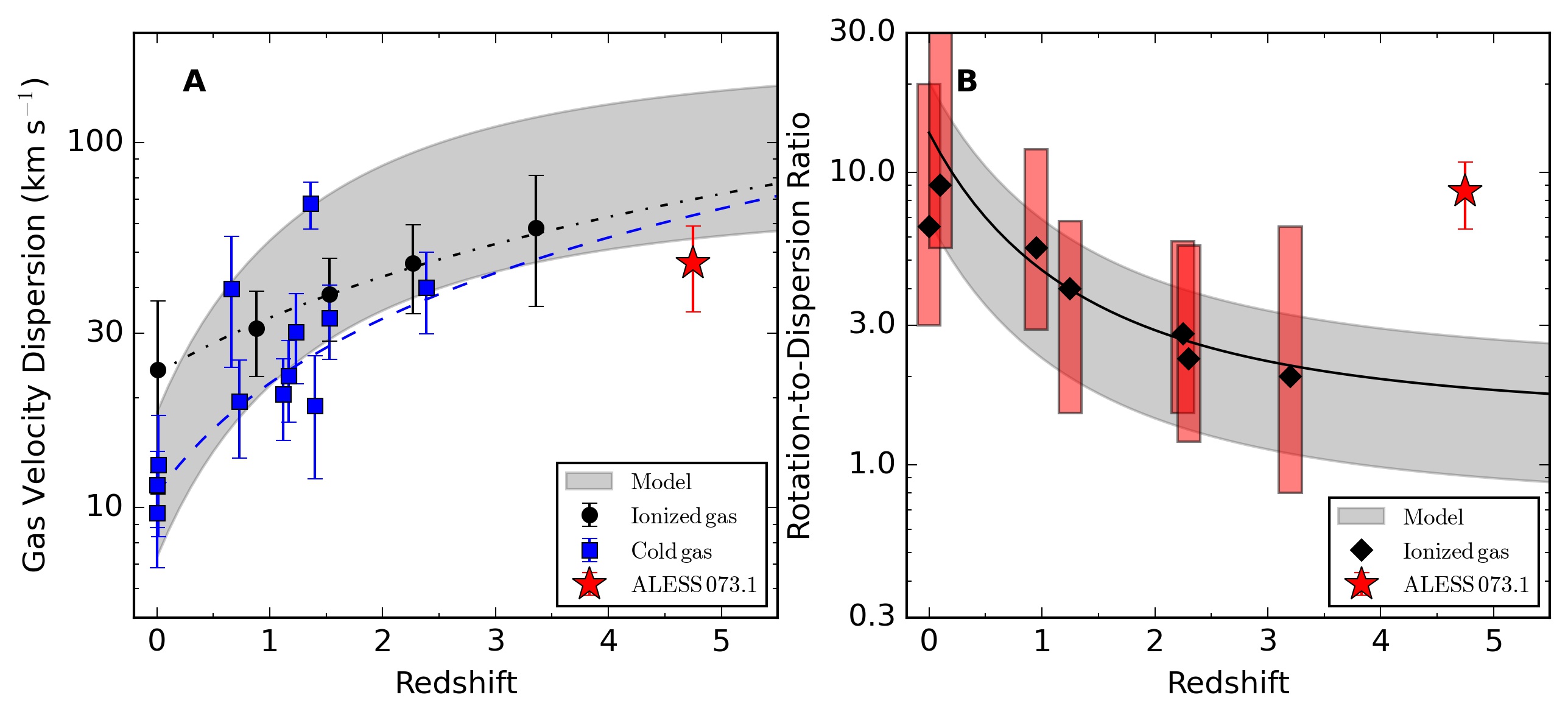}
\caption{\textbf{Turbulence and rotational support of galaxies as a function of cosmic time.} \textbf{(A)} Gas velocity dispersion and \textbf{(B)} $V_{\rm rot}/\sigma_{\rm V}$ versus redshift. Grey bands show a semi-empirical model based on Toomre's disk instability parameter \cite{Wisnioski2015}. In panel A, the black dot-dashed and the blue dashed lines show the velocity dispersion evolution inferred from warm ionized gas data (black circles; averages from galaxy samples) and cold neutral gas data (blue squares; averages for $z=0$ and individual galaxies for $z>0.5$), respectively \cite{Ubler2019}. In panel B, black diamonds and red bars show, respectively, the median and 90$\%$ distribution of ionized-gas surveys \cite{Wisnioski2015}. The red star represents ALESS\,073.1. Error bars correspond to $\pm$1$\sigma$ uncertainties.}
\end{figure*}

Gas disks in the early Universe are expected to be more turbulent than their local analogues: the gas velocity dispersion is thought to increase with redshift, whereas the degree of rotational support ($V_{\rm rot}/\sigma_{V}$) decreases \cite{Wisnioski2015, Ubler2019}. The enhanced turbulence may be driven by gravitational instabilities from cold gas flows, and/or by stellar and AGN feedback \cite{Lehnert2009}. At $z\geq1$, however, the limited spatial resolution can bias the measurements of $V_{\rm rot}$ and $\sigma_{V}$ (beam smearing effects), leading to systematic overestimates of $\sigma_{V}$ and underestimates of $V_{\rm rot}/\sigma_{V}$ \cite{DiTeodoro2016}. The gas velocity dispersion of ALESS\,073.1 is consistent with the extrapolations from data at $z<3.5$ (Fig. 3A) but lies at the low $\sigma_{\rm V}$ boundary of the prediction of a semi-empirical model based on Toomre's disk instability parameter \cite{Wisnioski2015}, despite the galaxy hosts a starburst and a dust-obscured AGN. The rotational support of the gas disk of ALESS\,073.1 is much higher than the value predicted by the same model (Fig. 3B), which has been used to describe kinematic measurements at $z<3.5$ \cite{Wisnioski2015}. The $V_{\rm rot}/\sigma_{V}$-ratio is similar to that of gas disks in the local Universe (Fig. 3B). This suggests that both starburst and AGN feedback have only a gentle effect on the cold interstellar medium of ALESS\,073.1 because the \cii\ disk is regular and unperturbed, contrary to galaxy formation models with violent feedback.

It remains unknown whether bulges exist in the early Universe ($z>4$) given the lack of high-resolution infrared images probing the stellar mass distribution. In ALESS\,073.1 we use the gas kinematics to infer the presence of a central massive component that is not traced by the dust. This is likely a stellar bulge hosting a super-massive black hole. Bulges form either from the secular evolution of stellar disks or after major galaxy mergers \cite{Kormendy2004}. Both mechanisms could be at play in ALESS\,073.1, but they must act on short timescales because the Universe was only 1.2 billion years old at $z\simeq4.75$. Any major merger must have happened at even earlier cosmic epochs to provide time for the gas kinematics to relax to regular rotation. The orbital time at the last measured point of the rotation curve is about $5\times10^7$ years, and about five revolutions may be required to relax the disk, so any major merger must have happened at $z>5.5$. ALMA observations of a quasar-host galaxy at $z\simeq6.6$ with similar spatial resolution show that the \cii\ kinematics can be heavily disturbed at this younger epoch ($\sim$0.8 billion years after the Big Bang), possibly owing to mergers or interactions \cite{Venemans2019}. Bulges may also develop from star formation inside AGN-driven gas outflows \cite{Ishibashi2014, Maiolino2017}, which could occur on short timescales. Although we observe only a single object, we conclude that the Universe produced regularly rotating galaxy disks with prominent bulges at $<$10$\%$ of its current age. This implies that the formation of massive galaxies and their central bulges must be a fast and efficient process.

\section*{Acknowledgments}
We thank L. Legrand and H. \"Ubler for providing tabular data from their publications. F.L. thanks P. Li for technical support with the Markov-Chain-Monte-Carlo fitting code. 

\paragraph{Funding:} F.L. was supported by the ESO fellowship during the initial stages of this project. E.M.D.T. was supported by the U.S. National Science Foundation under grant 1616177. F.F. acknowledges support from the Friedrich Wilhelm Bessel Research Award Programme of the Alexander von Humboldt Foundation. A.W.S.M. is supported by a Dunlap Fellowship at the Dunlap Institute for Astronomy \& Astrophysics, funded through an endowment established by the David Dunlap family and the University of Toronto. R.M. acknowledges ERC Advanced Grant 695671 ``QUENCH'' and support from the Science and Technology Facilities Council (STFC).

\paragraph{Authors contributions:} F.L.: conceptualization, data curation, formal analysis, investigation, methodology, project administration, software, validation, visualization, writing - original draft. E.M.D.T.: conceptualization, formal analysis, methodology, software, visualization, writing - review \& editing. F.F.: conceptualization, methodology, software, writing - review \& editing. A.W.S.M.: formal analysis, writing - review \& editing. Z.-Y.Z.: formal analysis, writing - review \& editing. CdB: data curation, writing - review \& editing. T.A.D.: resources, writing - review \& editing. R.M.: resources, writing - review \& editing.

\paragraph{Competing interests:} The authors have no competing interests.

\paragraph{Data and materials availability:} This paper makes use of the following ALMA data:\\ ADS/JAO.ALMA \#2017.1.01471.S, \#2015.1.00456.S, and \#2011.0.00124.S. ALMA is a partnership of ESO (representing its member states), NSF (USA) and NINS (Japan), together with NRC (Canada), MOST and ASIAA (Taiwan), and KASI (Republic of Korea), in cooperation with the Republic of Chile. The Joint ALMA Observatory is operated by ESO, AUI/NRAO and NAOJ. The data are available at the ALMA archive (https://almascience.eso.org/asax/). The paper also makes us of Hubble Space Telescope data from proposal ID 12061 available at https://archive.stsci.edu/hst/ (dataset ibeuglqlq). The observed \cii\ data cube (Data S1), the \cii\ moment maps (Data S2 and S3), the model outputs from \textsc{$^{\rm 3D}$Barolo} (Data S4-S6), the model outputs from \textsc{Galfit} (Data S7), and the best-fitting mass model (Data S8) are available as supplementary data files.

\renewcommand{\thefigure}{S\arabic{figure}}
\setcounter{figure}{0}
\renewcommand{\theequation}{S\arabic{equation}}
\renewcommand{\thetable}{S\arabic{table}}

\section*{Materials and Methods} 

\paragraph*{Observations and data reduction.}

ALESS\,073.1 was observed on 18 October 2018 with ALMA Band-7 and 47 working antennas, giving minimum and maximum baselines of 15 and 2517 m, respectively. The weather conditions were good with precipitable water vapor between 0.35 and 0.40 mm. The flux, bandpass, and pointing calibrator was ALMA J0522-3627 (B\,0521-365), while the phase calibrator was ALMA J0348-2749 (D\,0346-279). A high-resolution spectral window with a bandwidth of 1.875 GHz was covered with 480 channels and centered at 329.638 GHz to target the redshifted \cii\ emission line at a rest-frame frequency of 1900.539 GHz. Three low-resolution spectral windows with bandwidths of 2 GHz were covered with 128 channels each and centered at 331.201, 317.280, and 319.156 GHz to target the dust-emitting continuum. Two execution blocks of about 45 minutes each gave a total on-source integration time of 1.5 hours. Two additional execution blocks of 45 minutes were flagged as ``semi-pass'' by the ALMA quality-assurance team. We attempted to calibrate these additional data but concluded that they were unavoidably damaged due to time-dependent phase jumps. We also analysed archival data of ALESS\,073.1 from 2012 \cite{DeBreuck2014} and 2015 \cite{Gullberg2018}. The 2012 data were combined with the 2018 data, increasing the overall signal-to-noise ratio with their short baselines between 18 and 402 m. We used the 2015 data for the continuum imaging only, because the \cii\ emission was filtered out by the very long baselines (up to 16.2 km) of those observations \cite{Gullberg2018}.

The data reduction was performed using the Common Astronomy Software Applications (\textsc{Casa}) package \cite{McMullin2007}. The Fourier-plane data were flagged and calibrated using the standard \textsc{Casa} pipeline. Data from different ALMA cycles were taken with different spectral setups and definition of visibility weights. We first split the science target from the Fourier-plane data and used the \textsc{Casa} task \texttt{oldstatwt} on the line-free channels to set consistent visibility weights among the different datasets. Next, we regridded the line-data to a common bandwidth of about 1.6 GHz using 102 channels with a spectral width of 15.626 MHz, corresponding to a velocity width of $\sim$14 km s$^{-1}$.

The combined dataset was imaged using a H\"ogbom deconvolver with Briggs' robust parameter of 0.5 \cite{Briggs1995}. A continuum image was obtained combining all four spectral windows, excluding channels with line emission, reaching a sensitivity of 18 $\upmu$Jy beam$^{-1}$ (where 1 $\upmu$Jy = 10$^{-32}$ W m$^{-2}$ Hz$^{-1}$). The synthesized beam of the continuum map is $0.123'' \times 0.095''$ with a position angle (PA) of $-63.9^{\circ}$. A dirty \cii\ cube was obtained without any continuum subtraction and cleaning: these reduction steps were performed in the image plane using the Groningen Imaging Processing System (\textsc{Gipsy}) package \cite{Vogelaar2001}. The synthesized beam of the \cii\ cube is $0.129''\times 0.105''$ with a PA of $-63.6$. The root-mean square (rms) noise of the dirty \cii\ cube depends on frequency because a different number of visibilities from different ALMA cycles are combined in each rebinned channel. This is taken into account during the cleaning process with \textsc{Gipsy}.

The continuum was subtracted from the \cii\ cube in the image plane, after fitting a first-order polynomial to the line-free channels. The cleaning was performed within a Boolean mask that follows the kinematic structure of the \cii\ emission. The Boolean mask was constructed by smoothing the continuum-subtracted \cii\ cube to a resolution of 0.4$''$ and clipping at 3$\sigma_{\rm s}$, where $\sigma_{\rm s}$ is the rms noise of the smoothed cube. The cleaning was done down to 3$\sigma_{\rm d}$, where $\sigma_{\rm d}$ is the rms noise of the continuum-subtracted dirty cube. The cleaned cube was spectrally smoothed using the Hanning scheme, giving a velocity resolution of $\sim$28 km s$^{-1}$ (twice the channel width). This increased the signal-to-noise ratio, giving a mean rms noise of 0.18 mJy beam$^{-1}$ per channel.

\begin{figure*}
\centering
\includegraphics[width=16cm]{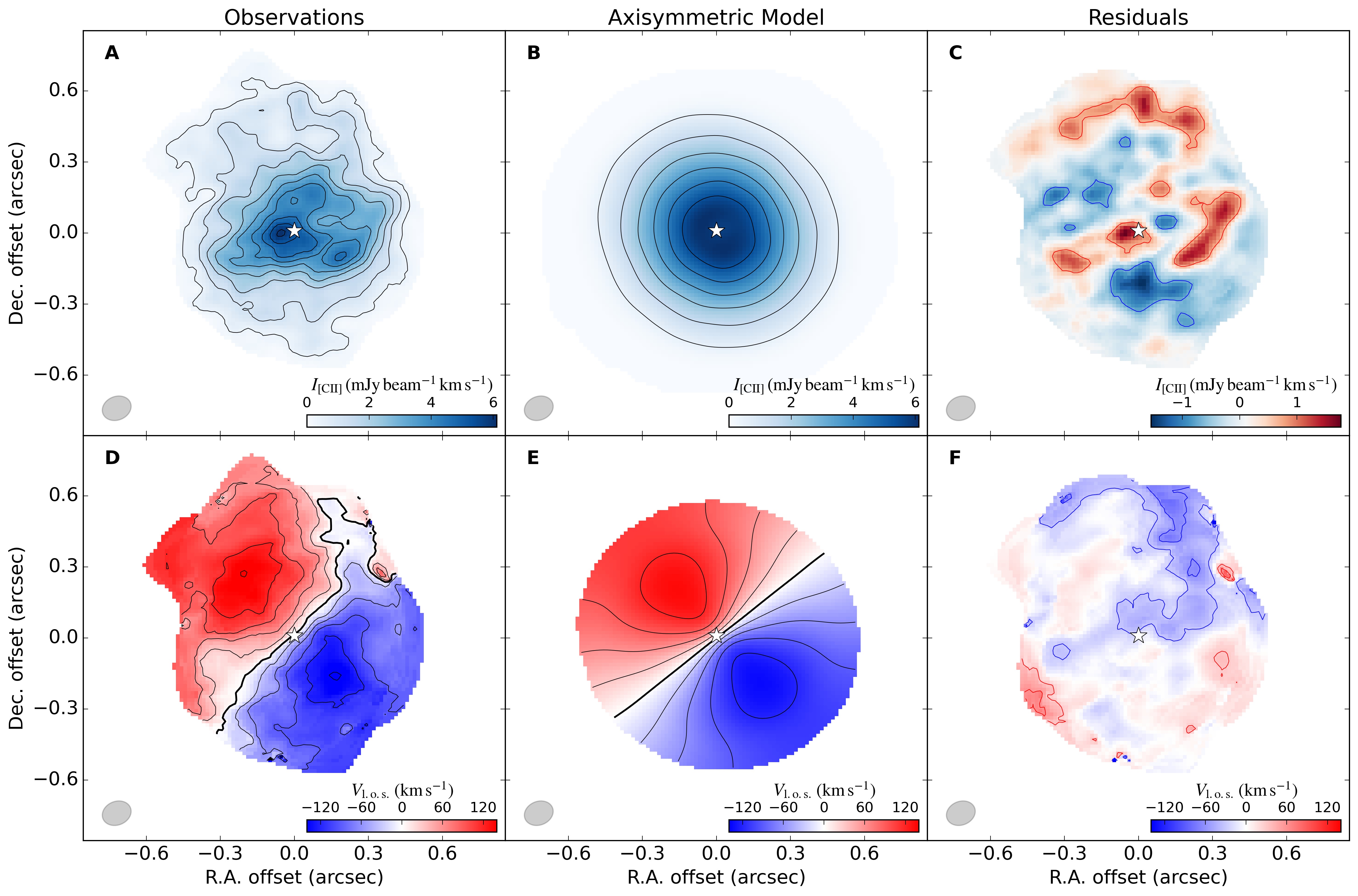}
\caption{\textbf{Moment Maps:} integrated \cii\ intensity ($I_{\rm [CII]}$) map from the observed ({\bf A}), model ({\bf B}), and residual data cube ({\bf C}); \cii\ line-of-sight velocity ($V_{\rm l.o.s.}$) map from the observed ({\bf D}), model ({\bf E}), and residual data cube ({\bf F}). North is up and east is left. The kinematic centre (R.A. = 03$^{\rm h}$ 32$^{\rm m}$ 29.295$^{\rm s}$; Dec. = $-$27$^\circ$ 56$'$ 19.60$''$) is shown with a white star. The beam size is plotted as the grey ellipse in the bottom-left corner. In panels A and B, iso-emission contours range from 0.35 mJy km\,s$^{-1}$ ($\sim$3$\sigma$) to 6 mJy km s$^{-1}$ in steps of 0.7 mJy km s$^{-1}$. In panel C, residual contours are at $-$6$\sigma$ (blue) and $+$6$\sigma$ (red); non-axisymmetric, arm-like features are detected. In panels D and E, iso-velocity contours range from $-120$ to $+120$ km\,s$^{-1}$ in steps of 30 km\,s$^{-1}$; the bold contour highlights the systemic velocity (set to zero). In panel F, the residual contours are at $-60$ and $-30$ km\,s$^{-1}$ (blue) as well as $+30$ and $+60$ km\,s$^{-1}$ (red). Non-circular motions above the velocity resolution ($\sim$30 km\,s$^{-1}$) are detected along an arm-like feature towards the northwest (see also Fig. S4).}
\end{figure*}

\paragraph*{Gas distribution and kinematics.}

We assume a $\Lambda$CDM cosmology adopting the Planck 2018 cosmological parameters  \cite{Planck2018}: matter density $\Omega_{\rm m} = 0.315$, cosmological constant density $\Omega_{\Lambda} = 0.685$, and Hubble constant $H_0 = 67.4$ km s$^{-1}$ Mpc$^{-1}$. In this cosmology, 0.1$''$ corresponds to 657.7 pc at the redshift of ALESS\,073.1 ($z=4.7555$). Coordinates of Right Ascension (R.A.) and Declination (Dec.) are given using the International Celestial Reference System (ICRS).

The \cii\ cube was analysed using the \textsc{$^{\rm 3D}$Barolo} package \cite{DiTeodoro2015}. Moment maps were constructed using the ``smooth \& search'' option, setting the \texttt{threshVelocity} parameter to 1. A pseudo-3$\sigma$ surface brightness contour was estimated following standard procedures \cite{Lelli2014}. The \cii\ intensity map (moment zero) and velocity field (moment one) are shown in Figure S1 (A and D). The \cii\ line is broadened by unresolved differences in rotation within the resolution element (the beam-smearing effect). \textsc{$^{\rm 3D}$Barolo} directly models the 3D cube taking beam-smearing effects into account, so does not rely on 2D moment maps.

\textsc{$^{\rm 3D}$Barolo} divides the galaxy into a series of rings with each ring described by nine parameters: coordinates of the kinematic center ($x_0, y_0$), position angle (PA), inclination ($i$), vertical thickness ($z_0$), systemic velocity ($V_{\rm sys}$), rotation velocity ($V_{\rm rot}$), radial velocity ($V_{\rm rad}$), and velocity dispersion ($\sigma_{\rm V}$). We use nine rings with a width of 0.06$''$, corresponding to half of the resolution element $\sqrt{\theta_{\rm x}\times \theta_{\rm y}}$, where $\theta_{\rm x}$ and $\theta_{\rm y}$ are the major and minor axes of the synthesized beam. For simplicity, we adopt a fully axisymmetric disk model, so the surface density of each ring is directly computed from the observed \cii\ intensity map using azimuthal averages. \textsc{$^{\rm 3D}$Barolo} can also produce disk models with non-axisymmetric gas distributions by renormalizing the flux density of each spatial pixel to the observed intensity map. This technique was introduced to account for rotating disks with massive gas clumps and/or feedback-driven cavities \cite{Lelli2012a, Lelli2012b} but it appears unnecessary for ALESS\,073.1. The non-axisymmetric approach would be nearly insensitive to the inclination angle because the observed intensity map (moment zero) is always reproduced by construction. We therefore stick to the axisymmetric model in which the gas surface density depends on radius ($R$) only.

We fix the vertical thickness of the gas disk to 300 pc; the value of $z_0$ has little effect on our results because the disk is nearly face-on. The radial velocity is negligible because the kinematic major axis of the disk is perpendicular to the kinematic minor axis (Fig. S1D), so we fix it to zero. If we leave $V_{\rm rad}$ as a free parameter, we find values consistent with zero within the uncertainties, so we conclude that large-scale radial motions in the disk must be smaller than $\sim$30 km\,s$^{-1}$ (approximately the velocity resolution).

A first run of \textsc{$^{\rm 3D}$Barolo} is used to estimate ($x_0, y_0$), PA, and $V_{\rm sys}$, considering the arithmetic average of the best-fitting values across the rings. When all parameters are left free, \textsc{$^{\rm 3D}$Barolo} occasionally converges to local rather than global minima giving unphysical solutions. We verified that the best-fitting values of ($x_0, y_0$), $V_{\rm sys}$, and PA are plausible by visually inspecting \cii\ channel maps, position-velocity (PV) diagrams, and moment maps. The PA is then fixed to 40$^{\circ}$ because the velocity field does not show any strong indication for a warped disk within $R\simeq0.45''$. At larger radii, the asymmetric \cii\ extension to the north may potentially be due to a warp: this would only affect the outermost 2-3 points of the rotation curve, so we keep the PA constant at all radii rather than modeling an uncertain, asymmetric warp. This anomalous kinematic component corresponds to non-circular motions of about 30-60 km\,s$^{-1}$ along an extended arm-like feature (Figs. S1 and S4). Our conclusion of a central massive bulge is driven by the rotation velocities at small radii, where there is no evidence for a warped disk, nor for non-circular motions.

A second run of \textsc{$^{\rm 3D}$Barolo} is used to estimate an average inclination of $25^{\circ}\pm3^{\circ}$; the uncertainty is taken to be the standard deviation across the rings. This value of $i$ is confirmed by comparing the model \cii\ intensity map with the observed one, but its formal uncertainty should be taken with caution because the \cii\ disk is close to face-on. The inclination angle has a large effect on the normalization of the best-fitting rotation curve, so we build mass models treating $i$ as a free parameter in a Bayesian context, scaling the rotation velocities by $\sin(i)$ and using the value of $i=25^{\circ}\pm3^{\circ}$ as a Gaussian prior. A previous kinematic analysis of ALESS\,073.1 using the ALMA 2012 data (with a lower spatial resolution of about 0.5$''$) gave different inclination values depending on the fitting technique and modeling assumptions \cite{DeBreuck2014}: $53^{\circ}\pm9^{\circ}$ fitting 2D maps with an exponential disk model, $50^{\circ}\pm8^{\circ}$ fitting 2D maps with an arctangent model, and $29^{\circ} \pm 4^{\circ}$ fitting the 3D cube with the \textsc{Kinms} software \cite{Davis2013}. Our 2018 observations resolve the \cii\ distribution and kinematics, so we can directly infer the disk inclination without a-priori assumptions on the gas density profile and rotation curve shape. The resulting inclination of $25^{\circ}\pm3^{\circ}$ is fully consistent with the previous 3D determination, but is discordant at the 2.5$\sigma$ level with the 2D determinations. When analysing poorly resolved data, it is preferable to fit the 3D cube over the 2D moment maps \cite{DiTeodoro2015}. For example, the asymmetric \cii\ extension to the north of ALESS\,073.1 may mimic a more inclined disk in moment maps at lower spatial resolutions of $\sim$0.5$''$ (as in the ALMA 2012 data), leading the previous 2D fitting to infer higher inclinations.

\begin{figure*}
\centering
\includegraphics[width=16cm]{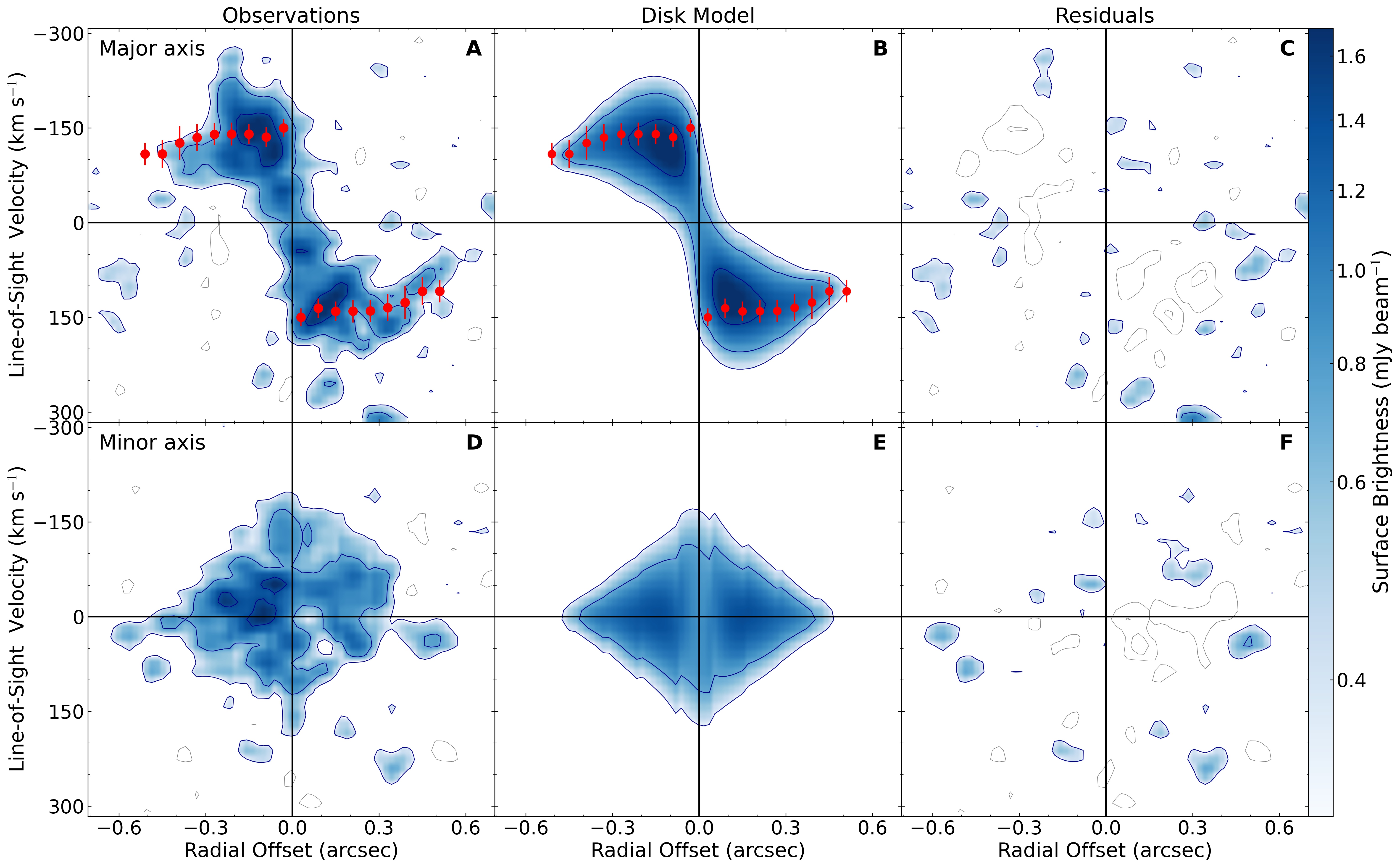}
\caption{\textbf{Kinematic Modeling:} position-velocity diagrams along the disk major axis from the observed ({\textbf A}), model ({\textbf B}), and residual ({\textbf C}) data cube; position-velocity diagrams along the disk minor axis from the observed ({\textbf D}), model ({\textbf E}), and residual data cube ({\textbf F}). Horizontal and vertical lines correspond to the galaxy systemic velocity and kinematic center, respectively. Blue contours are at 2$\sigma$, 4$\sigma$, and 8$\sigma$; grey contours are at -2$\sigma$ and -4$\sigma$. In panels A and B, red dots with $\pm$1$\sigma$ error bars show the projected rotation curve $V_{\rm rot}\sin(i)$.}
\end{figure*}

The final run of \textsc{$^{\rm 3D}$Barolo} uses only $V_{\rm rot}$ and $\sigma_{\rm V}$ as free parameters. For each parameter, \textsc{$^{\rm 3D}$Barolo} provides asymmetric uncertainties ($\delta_{+}$, $\delta_{-}$) corresponding to a variation of 5$\%$ of the residuals from the global minimum. We consider these errors as 2$\sigma$ deviations from the best-fitting value and compute symmetric 1$\sigma$ uncertainties as $\sqrt{\delta_{+}\delta_{-}}/2$, which are used in our Bayesian mass models (see Eq. 1). Panels B and E of Fig. S1 show moment-zero and moment-one maps from our best-fitting 3D model, while panels C and F show the residual maps. An axisymmetric, rotating disk model provides a good match to the data. The residual intensity map shows significant ($>6\sigma$) and spatially resolved ($>1$ beam) non-axisymmetric structures with an arm-like morphology. The non-axisymmetric structures to the south are also seen in the dust distribution, as we discuss below (see Fig. S4). The residual velocity field shows that non-circular motions are generally negligible (below our resolution limit of $\sim$30 km\,s$^{-1}$), except in the anomalous extension to the northwest. This may be due to recently accreted gas, streaming motions along spiral arms, or to a warped gas disk. The low signal-to-noise ratio in the outer regions prevents us from determining which.

To assess the reliability of our kinematic model, Figure S2 shows PV diagrams along the major and minor axis of the disk. PV diagrams are a direct representation of the data because they do not require masking or clipping, unlike moment maps. The PV diagrams from the model cube reproduce the observations along both kinematic axes. The majority of the residuals are within $\pm2\sigma$, implying that any non-circular motion must be smaller than the velocity resolution ($<30$ km\,s$^{-1}$), or localized to spatial scales smaller than the ALMA beam ($<$0.7 kpc), or involve gas at lower column densities ($<$40 M$_\odot$ pc$^{-2}$). Inspection of the contours on the residual PV diagrams reveal three small features at $-4\sigma$ along the major axis and one small feature at $-4\sigma$ along the minor axis. The size of these features is similar to our resolution element (both in space and velocity), thus they are consistent with being noise peaks.

Figure S3 shows the \cii\ channel maps from the observed and model cubes, confirming that the overall kinematics is reproduced by a simple rotating disk model. The arm-like feature to the northwest is again an exception, visible at velocities between $+103$ and $-38$ km\,s$^{-1}$. Similar deviations due to non-circular motions are common in galaxies at $z=0$ \cite{Lelli2012a, Lelli2012b, Trachternach2008, Fraternali2001, Sancisi2008}, so do no represent some peculiar feature of high-$z$ galaxies.

\begin{figure*}
\centering
\includegraphics[width=14.3cm]{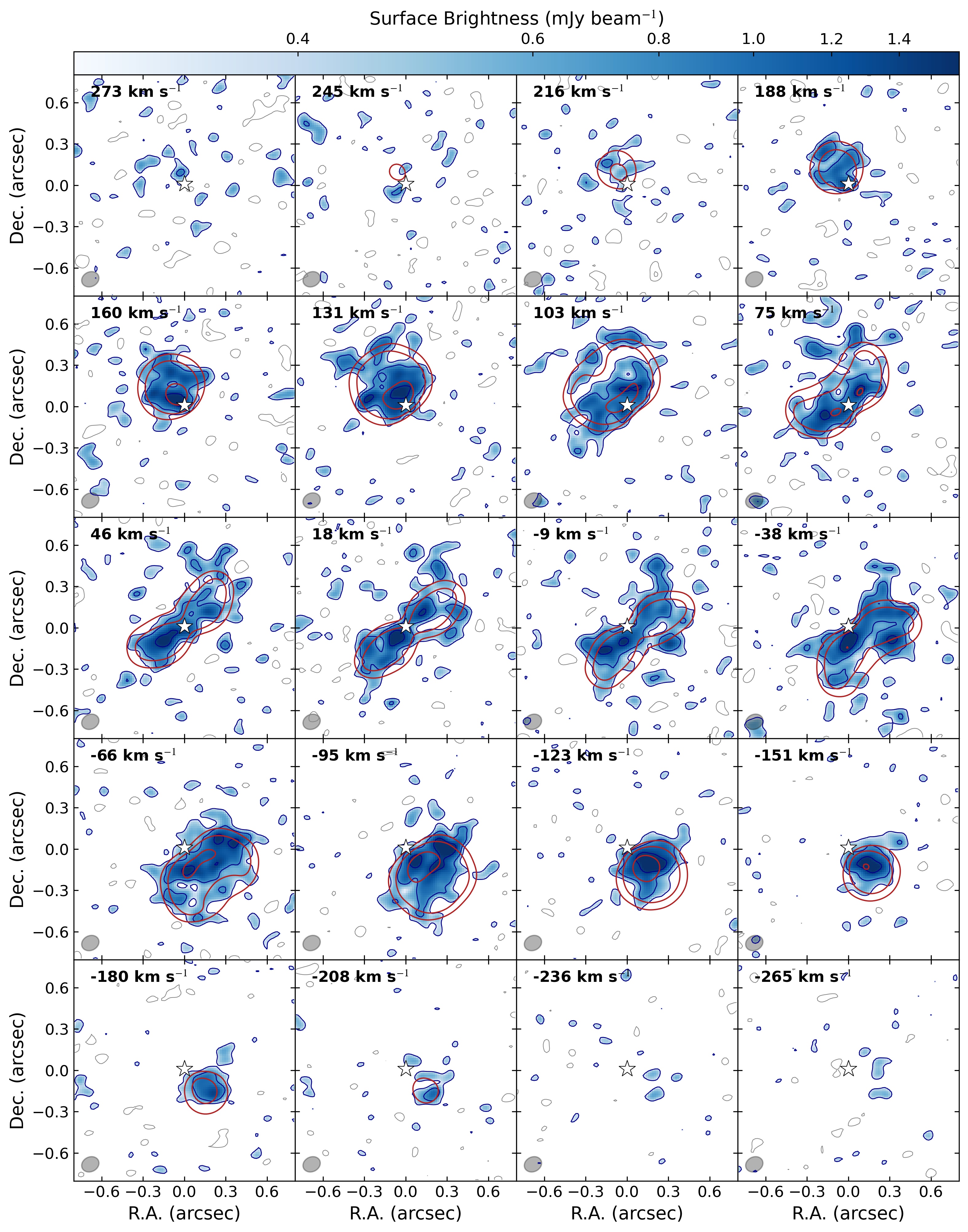}
\caption{\textbf{Channel Maps:} observed \cii\ emission (colorscale and blue contours) overlaid with the best-fitting disk model (red contours). Coloured contours are at 2$\sigma$, 4$\sigma$, 8$\sigma$. Grey contours are at -4$\sigma$ and -2$\sigma$. The beam size is plotted as the grey ellipse in the bottom-left corner. Each panel shows data at a different line-of-sight velocity, given in the top-left corner (with respect to the galaxy systemic velocity). The white star indicates the galaxy center.}
\end{figure*}

\paragraph*{Dust distribution.}

The dust distribution (Fig. S4A) appears elongated along a PA of about 110$^{\circ}$ which is different from the kinematic PA of about 40$^{\circ}$. The asymmetric shape of the ALMA beam, however, may play a role because the dust emission is more compact than the \cii\ emission, so it is less well resolved. Inspection of the continuum map shows non-axisymmetric structures to the south, distorting the iso-emission contours. We use the \textsc{Galfit} package \cite{Peng2010} to fit 2D analytic functions to the dust distribution taking the point-spread function (PSF) into account. Because the ALMA continuum map was cleaned down to the noise level, we consider a Gaussian PSF with a full-width half-maximum equal to the synthesized beam ($0.123'' \times 0.095''$ with PA of $-63.9^{\circ}$).

If we run \textsc{Galfit} leaving all parameters free, we do not find satisfactory results due to parameter degeneracies. Our goal is to check whether the dust orientation is consistent with the \cii\ orientation. Thus, we fix several \textsc{Galfit} parameters, adopting an exponential disk (S\'ersic index $n=1$) with the same center and inclination as the \cii\ disk (projected axis ratio of $0.9$). The free parameters are the intrinsic PA, the effective radius, and the integrated flux. Because the dust distribution has substructures, we use an iterative approach masking residual pixels above 5$\sigma$ until the fitting results converge \cite{Hodge2019}. \textsc{Galfit} returns $\rm{PA}=41^{\circ}\pm5^{\circ}$ at the first iteration and $\rm{PA}=36^{\circ}\pm6^{\circ}$ at the fifth, final iteration (Fig. S4B). These values are consistent with the kinematic PA. However, if we allow the axis ratio and/or the S\'ersic index to vary, a wide range of PA values are found, indicating that the intrinsic geometry cannot be constrained without any prior information. We conclude that the orientation of the dust disk is consistent with that of the \cii\ disk.

\begin{figure*}
\centering
\includegraphics[width=16cm]{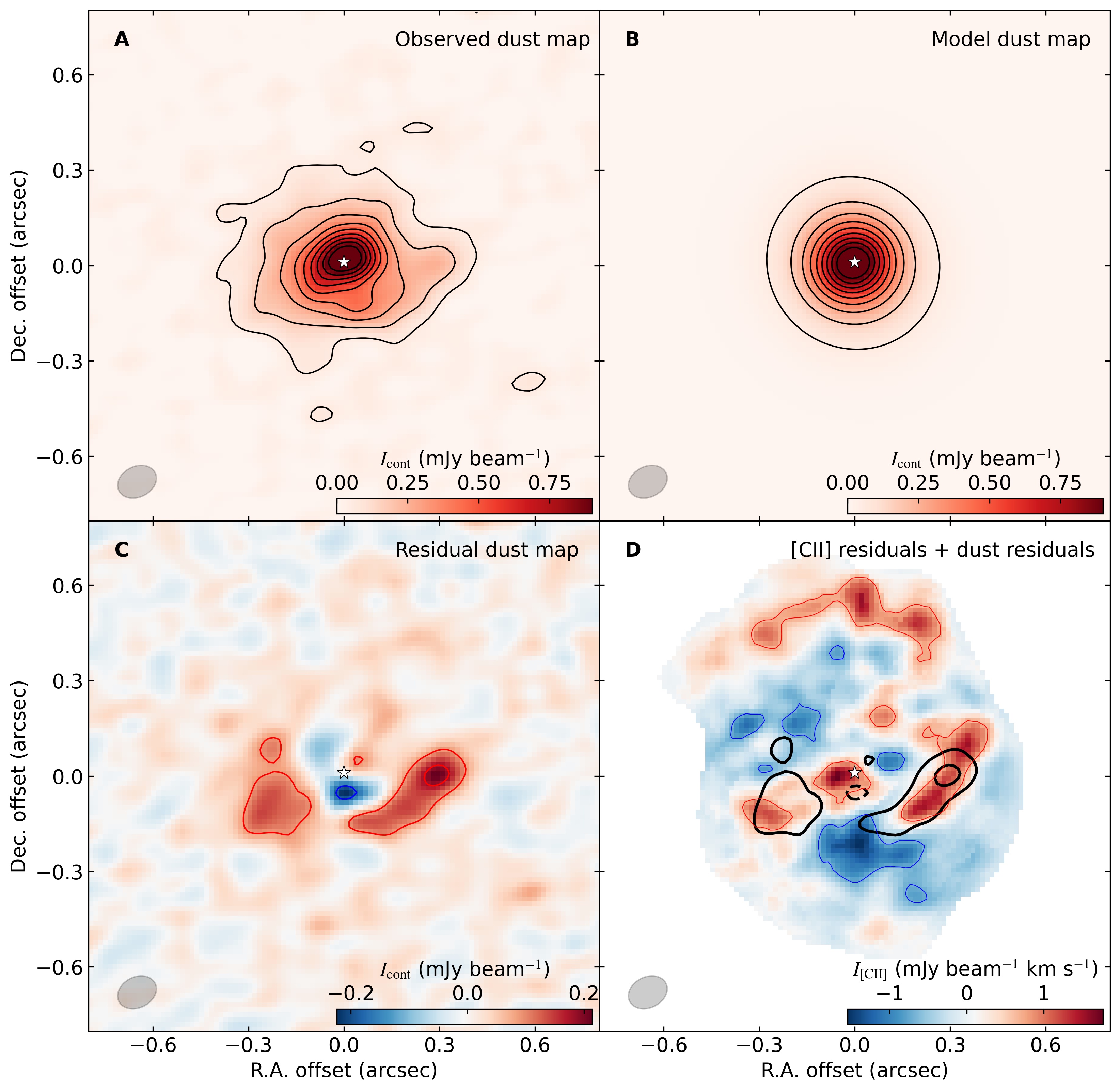}
\caption{\textbf{Dust Distribution:} observed continuum intensity ($I_{\rm cont}$) map at rest-frame 160 ${\upmu}$m (\textbf{A}), model dust map (\textbf{B}), residual dust map (\textbf{C}), and residual dust map overimposed on the \cii\ residual intensity map (\textbf{D}). North is up and east is left. The kinematic centre is shown with a white star. The beam size is plotted as the grey ellipse in the bottom-left corner. In panels A and B, iso-emission contours range from 0.055 ($\sim$3$\sigma$) to 1 mJy beam$^{-1}$ in steps of 0.11 mJ beam$^{-1}$. In panel C, residual contours are at $-$5$\sigma$ (blue), $+$5$\sigma$ and $+$10$\sigma$ (red); an arm-like feature is visible. In panel D, the same negative and positive contours are plotted with dashed and solid black lines, respectively; the residual \cii\ contours are at $-$6$\sigma$ (blue) and $+$6$\sigma$ (red); the dust and gaseous arm-like features to the southwest are approximately at the same location.}
\end{figure*}

The residual map (Fig. S4C) includes a prominent arm-like feature to the southwest which may connect to a broader feature to the southeast. Similar structures are also visible in the residual \cii\ intensity map (Fig. S4D). It is unlikely that these features are due to a disrupting merging companion or to a gas accretion event because the \cii\ kinematics appears largely unperturbed at these locations. We interpret these features as a spiral arm embedded in a rotating disk. Similar structures have been observed in high-resolution dust maps of other submillimeter galaxies at high redshifts \cite{Hodge2019}. Single-arm spiral structures can be excited by weak tidal perturbations due to, e.g., a nearby low-mass companion \cite{Athanassoula1978} or a past galaxy encounter \cite{Thomasson1989}. The \cii\ distribution displays a longer arm-like feature to the northwest (Fig. S4D), where the dust emission is weak or absent and the gas kinematics are less regular (Fig. S1). This may potentially represent a second arm.

\paragraph*{Mass Models.}

We build a mass model with four components: a cold gas disk, a stellar disk, a stellar bulge, and a DM halo. For each mass component, we compute the expected circular velocity of a test particle using the \texttt{Rotmod} task in \textsc{Gipsy}. We also present alternative models with fewer mass components, which have the advantage of having fewer free parameters but are less plausible within the standard galaxy-formation scenario (Figure S6). The fiducial mass model has six free parameters: disk inclination ($i$), gas mass ($M_{\rm gas}$), stellar disk mass ($M_{\rm disk}$), bulge mass ($M_{\rm bul}$), halo mass ($M_{200}$), and halo concentration ($C_{200}$).

The gravitational contribution of the cold gas is computed by solving Poisson's equation in cylindrical symmetry and considering a disk of finite thickness \cite{Casertano1983}. We use the azimuthally-averaged radial density profile from the observed \cii\ intensity map (Fig. 2B). This is more accurate than assuming simplified functional forms like exponential radial density profiles \cite{Sellwood1999}. For the vertical density distribution, we assume an exponential law with a fixed scale height of 300 pc \cite{vdK2011}. The vertical geometry has a minor effect on the resulting circular velocity in the disk mid-plane: the difference between a zero-thickness disk and the equivalent spherical distribution is at most 20$\%$ \cite{Noordermeer2008}. For numerical convenience, the gravitational contribution is normalized to an arbitrary mass of 10$^{10}$ M$_\odot$, which is then scaled up or down during the fitting process using a dimensionless parameter $\Upsilon_{\rm gas}$.

The gravitational contribution of the stellar disk is computed in the same way as for the gas disk. ALESS\,073.1 has been detected at multiple wavelengths but the stellar component has not been spatially resolved. We assume that the young stars are distributed in a similar way as the dust because the ultraviolet emission from young stars is absorbed by dust grains and re-emitted in the far-infrared. Dust grains at high redshifts, when the Universe was $\sim$1 Gyr old, are thought to be produced by asymptotic-giant-branch stars with masses between 3 and 8 M$_\odot$ after 10-100 Myr of stellar evolution and by core-collapse supernovae with lifetimes of about 10 Myr \cite{Gall2011}. The dynamical timescales ($R/V_{\rm rot}$) in ALESS\,073.1 are $\sim$50 Myr, so it is likely that dust and stars did not have enough time to alter their respective radial distributions. The azimuthally-averaged radial density profile from the ALMA continuum map can be described by an exponential model with a scale length of 700 pc (Fig. 2B). This scale length is similar to compact stellar disks in high-surface-brightness galaxies in the nearby Universe \cite{vdK2011}, although it is likely that the stellar disk of ALESS\,073.1 will grow in size from $z\simeq5$ to $z\simeq0$ following the inside-out galaxy formation scenario \cite{Pezzulli2015}. We use the dust radial density profile to solve Poisson's equation in cylindrical symmetry, adopting an exponential vertical distribution with scale-height of 300 pc. The gravitational contribution is again normalized to a mass of 10$^{10}$ M$_\odot$ and rescaled using a dimensionless parameter $\Upsilon_{\rm disk}$.

The gravitational contribution of the stellar bulge is computed assuming spherical symmetry \cite{Kent1986} and a De Vaucouleurs' projected density profile \cite{DeVauc1948}. To constrain the bulge effective radius ($R_{\rm e}$), we analyze an image of ALESS\,073.1 from the Hubble Space Telescope (HST) in the F160W filter \cite{Grogin2011}. We use \textsc{Galfit} \cite{Peng2010} and construct a PSF using seven nearby stars. The galaxy is unresolved, but we obtain an upper limit on $R_{\rm e}$ of about 300 pc (less than one HST pixel). The HST image may be affected by the AGN contribution because it measures the rest-frame near-ultraviolet emission, so the value of $R_{\rm e}$ is uncertain. Different values of $R_{\rm e}$ do not change our general result that there is a central bulge component, but slightly modify the best-fitting stellar masses. For example, a smaller value of $R_e = 200$ pc would decrease the bulge mass by about 20$\%$, while increasing the disk mass by a similar amount. In submillimeter galaxies at $z\simeq2-2.5$, the stellar component is sometimes observed to be more extended than the dust component \cite{Hodge2016, Lang2019}. If we assume that the stellar disk has a purely exponential radial profile with a scale-length larger than that of the dust ($>700$ pc), the stellar-disk contribution to the rotation curve would peak at larger radii, leaving a more massive bulge component in the central parts. Our fiducial mass model (assuming that the stellar disk is traced by the dust and the stellar bulge follows a De Vaucouleurs' profile) is conservative regarding the presence and mass of a central bulge component. Similarly to the gas and stellar disks, we introduce a dimensionless scaling parameter $\Upsilon_{\rm bul}$ normalized to a mass of 10$^{10}$ M$_\odot$.

The DM gravitational contribution is computed assuming a spherical halo with a Navarro-Frenk-White (NFW) density profile \cite{Navarro1996}. The halo mass $M_{200}$ is measured at $R_{200}$, the radius within which the mass volume density equals 200 times the critical density of the Universe. For convenience, we perform our model fitting by converting the halo mass to the circular speed at $R_{200}$, given by $V_{200} = [10\,G\,H(z)\,M_{200}]^{1/3}$ where $G$ is Newton's gravitational constant and $H(z)$ is the Hubble parameter. In our adopted cosmology, $H(z)$ is about 7.8 $H_0$ at $z=4.7555$. The halo concentration is defined as $C_{200} = R_{200}/R_h$, where $R_h$ is the characteristic scale-length of the NFW profile. 

\paragraph*{MCMC fitting.}

The best-fitting parameters are determined using a Markov-Chain-Monte-Carlo (MCMC) method in a Bayesian context. We define the likelihood $L = \exp({-0.5 \chi^2})$ with
\begin{equation}
 \chi^2 = \sum_{i}^{N} \frac{[V_{\rm obs}(R_i) - V_{\rm mod}(R_i)]^{2}}{\delta^{2}_{V_{\rm obs}}(R_i)},
\end{equation}
where $V_{\rm obs}$ is the observed rotation curve at radius $R_i$, $\delta_{V_{\rm obs}}$ is the associated error, and $V_{\rm mod}$ is the model rotation curve obtained by summing in quadrature the velocity contributions of each mass component, namely:
\begin{equation}
 V_{\rm mod}^2 = \Upsilon_{\rm gas}V_{\rm gas}^2 + \Upsilon_{\rm disk}V_{\rm disk}^2 + \Upsilon_{\rm bul}V_{\rm bul}^2 + V_{\rm halo}^2(C_{200}, V_{200}).
\end{equation}

We adopt a Gaussian prior on $i$ with a central value of 25$^{\circ}$ and a standard deviation of 3$^{\circ}$, taken from the modeling with \textsc{$^{\rm 3D}$Barolo}. The disk inclination changes the observed rotation velocities as $V_{\rm obs}\sin(i)$. For the other parameters, we adopt Gaussian or log-normal priors that are motivated by previous observations of ALESS\,073.1 and theoretical expectations, as we detail below. Gaussian and log-normal priors in Bayesian statistics ensure posterior propriety \cite{Tak2018}, but in our case they are also necessary because uninformative, flat priors would lead to parameter degeneracies. Our priors are summarized in Table S1. We impose boundaries to keep the parameters in a physical range: inclinations between 0 and 90 degrees; gas, disk and bulge masses between 10$^8$ and 10$^{12}$ M$_\odot$; halo velocity between 10 and 1000 km\,s$^{-1}$; and halo concentration between 0.01 and 100.

\begin{table}[thp]
\caption{MCMC priors.}
\label{tab:prop}
\centering
\begin{tabular}{l l c c}
\hline
Quantity                    & Prior Type & Center   & Standard Deviation \\
\hline
$i$ ($^{\circ}$)            & Gaussian   & 25.00    & 3.00 \\
$\log(\Upsilon_{\rm gas})$  & Lognormal  & 0.32     & 0.50 \\
$\log(\Upsilon_{\rm disk})$ & Lognormal  & 0.70     & 0.50 \\
$\log(\Upsilon_{\rm bul})$  & Lognormal  & 0.70     & 0.50 \\
$\log(V_{200})$             & Lognormal  & Eq.\,S3  & 0.45 \\
$C_{200}$                   & Gaussian   & 3.40     & 0.85 \\                                          
\hline
\end{tabular}
\end{table}

The cold gas mass ($M_{\rm gas}$) of ALESS\,073.1 is constrained by previous observations. The galaxy has been detected in the CO($J=2\rightarrow1$) line giving a molecular gas mass of $(1.6 \pm 0.3) \times 10^{10}$ M$_\odot$ \cite{Coppin2010}. This value assumes thermalized gas with $L_{{\rm CO}(J=2\rightarrow1)} = L_{{\rm CO}(J=1\rightarrow0)}$ and the CO-to-H$_2$ conversion factor $\alpha_{\rm CO} = 0.8$ M$_\odot$ (K km s$^{-1}$ pc$^{2}$)$^{-1}$, appropriate for starburst galaxies \cite{Tacconi2008}. The mass of atomic gas (hydrogen and helium) can be inferred from the CO($J=2\rightarrow1$) and \cii\ fluxes using models of photo-dissociation regions (PDR), giving a value of $(4.7 \pm 0.5)\times10^{9}$ M$_\odot$ \cite{DeBreuck2011, DeBreuck2014}. Thus, ALESS\,073.1 has an estimated cold gas mass (atomic plus molecular) of $2.1 \times 10^{10}$ M$_\odot$. Considering the large uncertainties in both the $\alpha_{\rm CO}$ factor and PDR modeling, we center the prior at $\log(\Upsilon_{\rm gas}) = 0.32$ with a standard deviation of 0.5 dex.

The total stellar mass ($M_\star$) of ALESS\,073.1 has been estimated by six different teams by fitting the spectral energy distribution (SED) with stellar population synthesis models \cite{Stark2007, Coppin2009, Wardlow2011, Gilli2014, Simpson2014, Wiklind2014}. ALESS\,073.1 has a well-sampled SED from the far-ultraviolet to the far-infrared, but these studies provide inconsistent estimates of $M_\star$ ranging from 5 to $13\times10^{10}$ M$_\odot$, possibly due to contributions from a starburst and AGN. For example, the AGN contribution in the optical portion of the SED can mimic an old stellar population. We adopt the arithmetic average of the six independent determinations ($10^{11}$ M$_\odot$) as our initial guess. This mass may be distributed between the disk and bulge components, so we center the disk prior at $\log(\Upsilon_{\rm disk}) = 0.7$ and the bulge prior at $\log(\Upsilon_{\rm bul}) = 0.7$, assuming a standard deviation of 0.5 dex for both components. Formally, $\Upsilon_{\rm bul}$ includes the mass of the central super-massive black hole, which cannot be independently measured with the available data. In nearby galaxies, the mass of the central black hole is typically 1$\%$ (or less) of the mass of the bulge \cite{Kormendy2013}. In high-$z$ galaxies the situation might be different, but super-massive black holes have been found to contribute $< 10\%$ of the central dynamical mass \cite{Pensabene2020}. Thus, the central black hole can have only a small effect on the measured $\Upsilon_{\rm bul}$.

To constrain the halo mass ($M_{200}$), we impose the stellar mass$-$halo mass relation from abundance$-$matching techniques, i.e., by matching the cumulative abundance of predicted DM halos to that of observed galaxies \cite{Vale2004, Guo2010}. The $M_\star-M_{200}$ relation is very uncertain in the early Universe. We adopt a simple log-linear relation:
\begin{equation}
\log(M_{200}) = 0.73\log(M_\star) + 4.61,
\end{equation}
which we derive from data at $z=4.5-5.5$ \cite{Legrand2019}. This relation assigns a galaxy with $M_\star = 5 \times 10^{10}$ M$_\odot$ a DM halo with $M_{200} = 2.6 \times 10^{12}$ M$_\odot$ (on average) giving a ``condensed'' baryon fraction of 0.019, which is about one order of magnitude lower than the cosmic value $\Omega_{\rm b}/\Omega_{\rm m} = 0.187$ \cite{Planck2018}. We assume a scatter of 0.45 dex in halo mass at fixed stellar mass \cite{Legrand2019}. The halo concentration ($C_{200}$) becomes nearly independent of halo mass at $z>3$, thus we assume a Gaussian prior centered at $C_{200} = 3.4$ with a standard deviation of 25$\%$ \cite{Dutton2014}.
Without imposing these two priors, the halo parameters would be unconstrained. The addition of a DM halo, however, is necessary to determine whether the observed rotation curve is consistent with the expectations of the $\Lambda$CDM cosmology.

The posterior probability distributions of the parameter set are mapped using the affine-invariant ensemble sampler \textsc{emcee} \cite{Foreman2013}. The MCMC chains are initialized with 200 walkers. Their starting positions are randomly assigned within these ranges for the fit parameters: $20^{\circ}<i<40^{\circ}$, $0 < \log(\Upsilon_{\rm gas}) < 1$, $0 < \log(\Upsilon_{\rm disk}) < 1$, $0 < \log(\Upsilon_{\rm bul}) < 1$, $2.4 < \log(V_{\rm 200}) < 2.6$, and $3.0 < C_{200} < 4.0$. The starting positions of the walkers have virtually no effect on the final results but help to ensure fast convergence of the chains. We run 1000 burn-in iterations, then the sampler is reset and run for another 2000 iterations. The \textsc{emcee} parameter \texttt{a}, which controls the size of the stretch move, is set equal to 2. This gives acceptance fractions around 50\%. The best fitting parameters are summarized in Table S2.

\begin{table}[thp]
\caption{MCMC results. The uncertainties correspond to 1$\sigma$ confidence regions (see Fig. S5).}
\label{tab:prop}
\centering
\begin{tabular}{l l}
\hline
$i$ ($^{\circ}$)         & 22.0$_{-2.9}^{+2.9}$ \\
$M_{\rm{gas}}$ (10$^{10}$ M$_{\odot}$) & 0.5$_{-0.4}^{+0.4}$ \\
$M_{\rm{disk}}$ (10$^{10}$ M$_{\odot}$)& 2.4$_{-1.0}^{+0.9}$ \\
$M_{\rm{bul}}$ (10$^{10}$ M$_{\odot}$) & 2.3$_{-0.6}^{+0.6}$\\
$V_{200}$ (km s$^{-1}$)                & 263$_{-80}^{+81}$\\
$C_{200}$                              & 2.9$_{-0.9}^{+0.9}$\\                                          
\hline
$M_{200}$ (10$^{11}$ M$_{\odot}$)      & 8.1$_{-2.7}^{+2.7}$\\
$M_{\rm{baryon}}$ (10$^{10}$ M$_{\odot}$) & 5.2$_{-1.2}^{+1.2}$\\
$M_{\rm{bul}}/M_{\rm{baryon}}$            & 0.44$_{-0.12}^{+0.11}$\\
\hline
\end{tabular}
\end{table}

\begin{figure*}
\centering
\includegraphics[width=16cm]{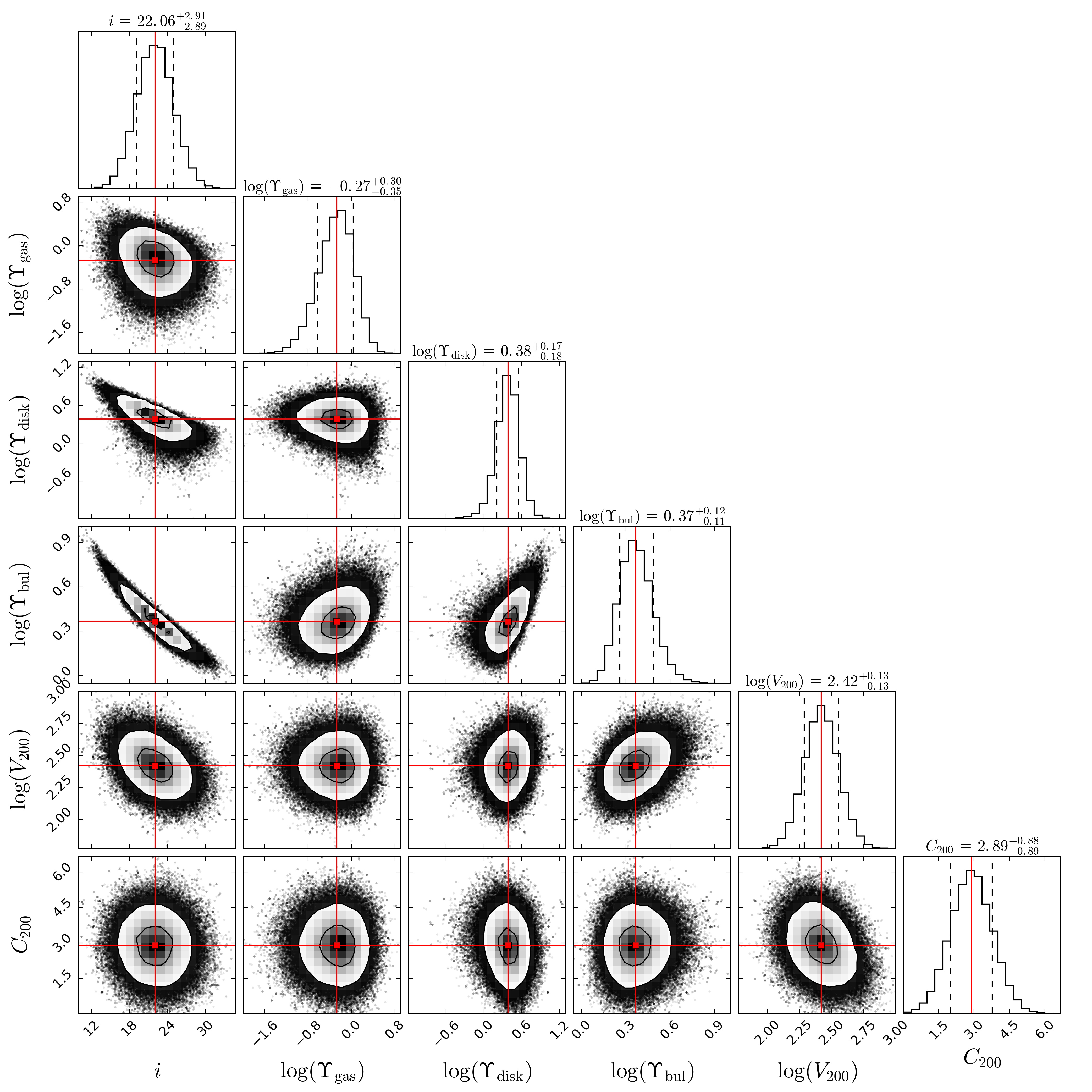}
\caption{\textbf{MCMC results:} the panels show the posterior probability distribution of pairs of mass model fitting parameters, and the marginalized probability distribution of each fitting parameter (histograms). Individual MCMC samples outside the 2$\sigma$ confidence region are shown with black dots, while binned MCMC samples inside the 2$\sigma$ confidence region are shown by the greyscale; the black contour corresponds to the 1$\sigma$ confidence region; the red squares and solid lines show median values. In the histograms, solid and dashed lines correspond to the median and $\pm$1$\sigma$ values, respectively. $\Upsilon_{\rm gas}$, $\Upsilon_{\rm disk}$ and $\Upsilon_{\rm bul}$ are normalized masses in units of 10$^{10}$ M$_\odot$. }
\end{figure*}

Figure S5 shows the posterior probability distributions of the fitting parameters of our fiducial mass model. We use \textsc{corner.py} \cite{Corner2016} and calculate 1$\sigma$ errors corresponding to the 68\% confidence interval of the marginalized 1-dimensional posterior probability distributions. As expected, both $M_{\rm disk}$ and $M_{\rm bul}$ are degenerate with $i$. This occurs because the overall normalization of the rotation curve is set by $1/\sin(i)$. For our inclination prior of $25^{\circ}\pm3^{\circ}$ (from the \textsc{$^{\rm 3D}$Barolo} models), the final best-fitting inclination is $22.0^{\circ}\pm2.9^{\circ}$, giving rotation velocities between 300 and 400 km\,s$^{-1}$ and pushing the baryonic masses towards the lower 1$\sigma$ boundary of the respective priors. Figure S5 shows that $M_{\rm disk}$ and $M_{\rm bul}$ are only weakly degenerate with each other, implying that the ratio $M_{\rm bul}/M_{\rm disk}$ is well constrained and nearly independent of $i$.

\paragraph*{Alternative Mass Models.}

We build more conservative mass models that have three gravitational components instead of four, excluding (in turn) the stellar bulge, the stellar disk, the gas disk, and the DM halo. Some of these models (e.g. the one without a DM halo) are statistically favored over our fiducial model because they provide a similarly accurate fit with a smaller number of free parameters. They are, however, less plausible within the standard $\Lambda$CDM paradigm of galaxy formation. For mass models without the stellar disk or stellar bulge, the Bayesian priors are revised such that the total fiducial stellar mass (10$^{11}$ M$_\odot$) is preserved.

\begin{figure*}
\centering
\includegraphics[width=7.8cm]{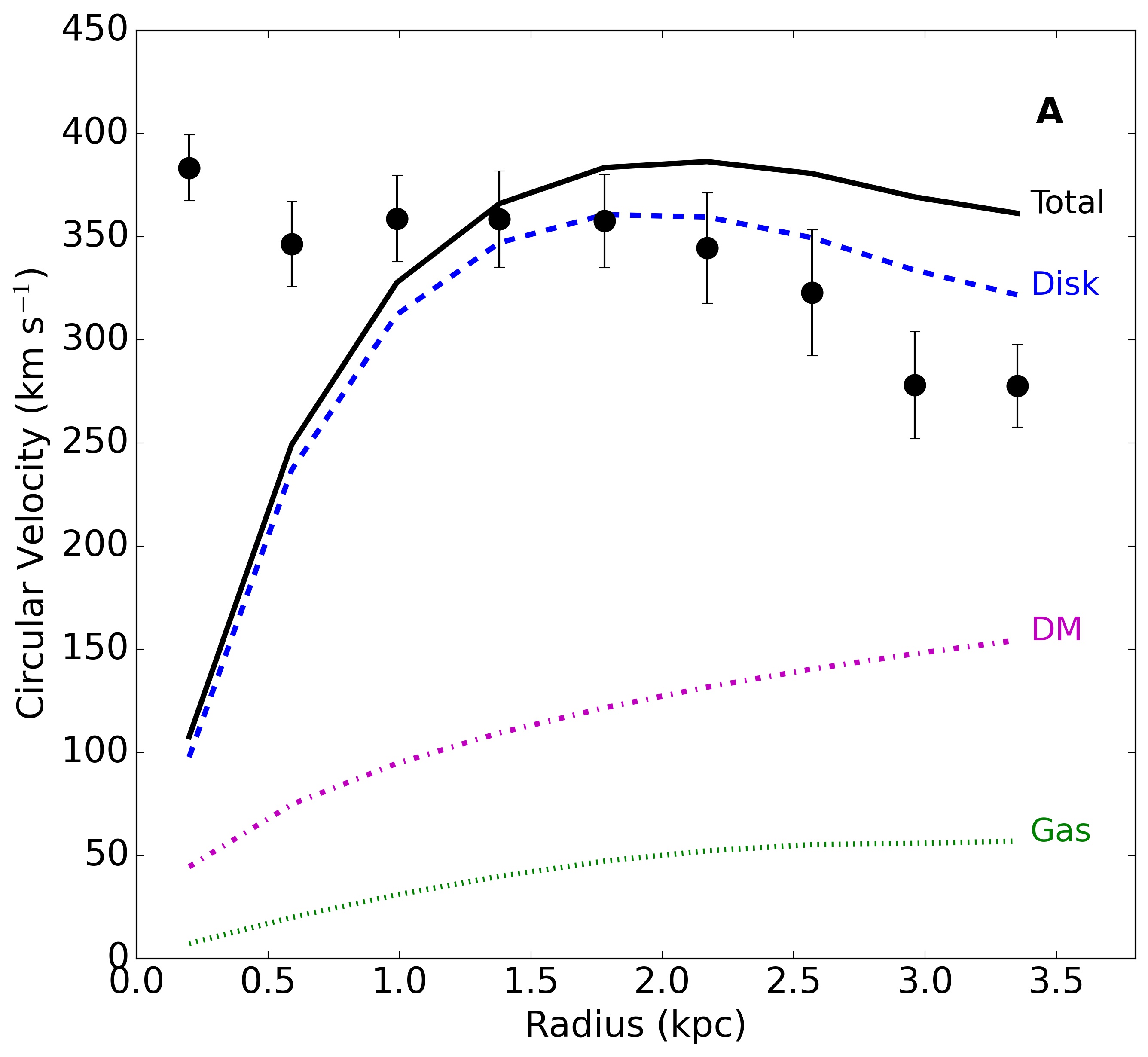}
\includegraphics[width=7.8cm]{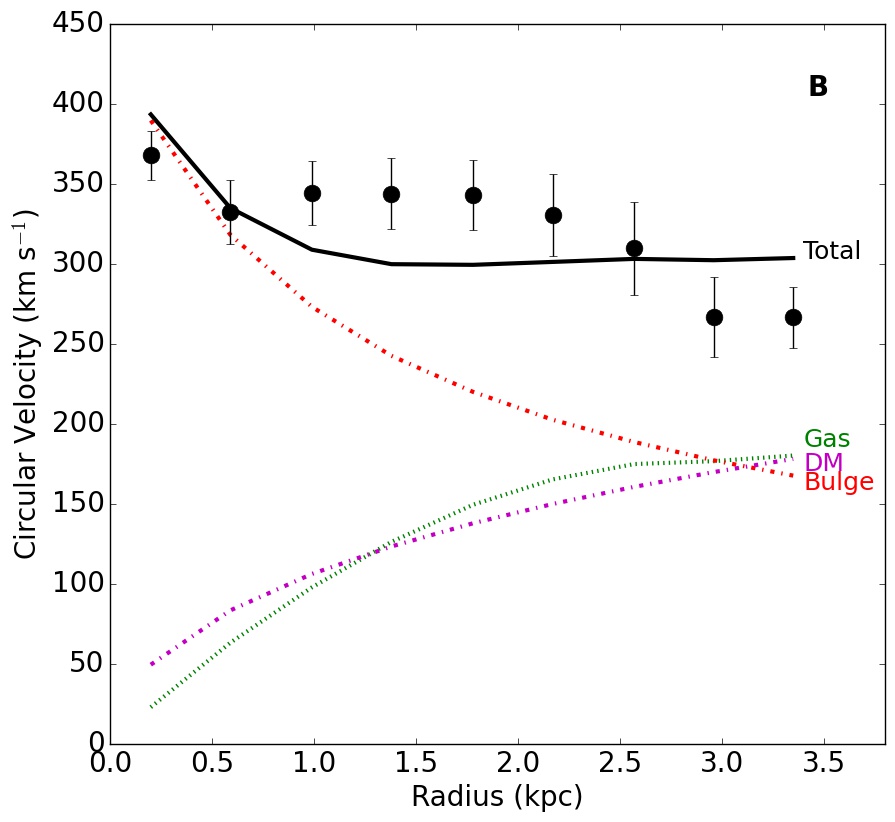}
\includegraphics[width=7.8cm]{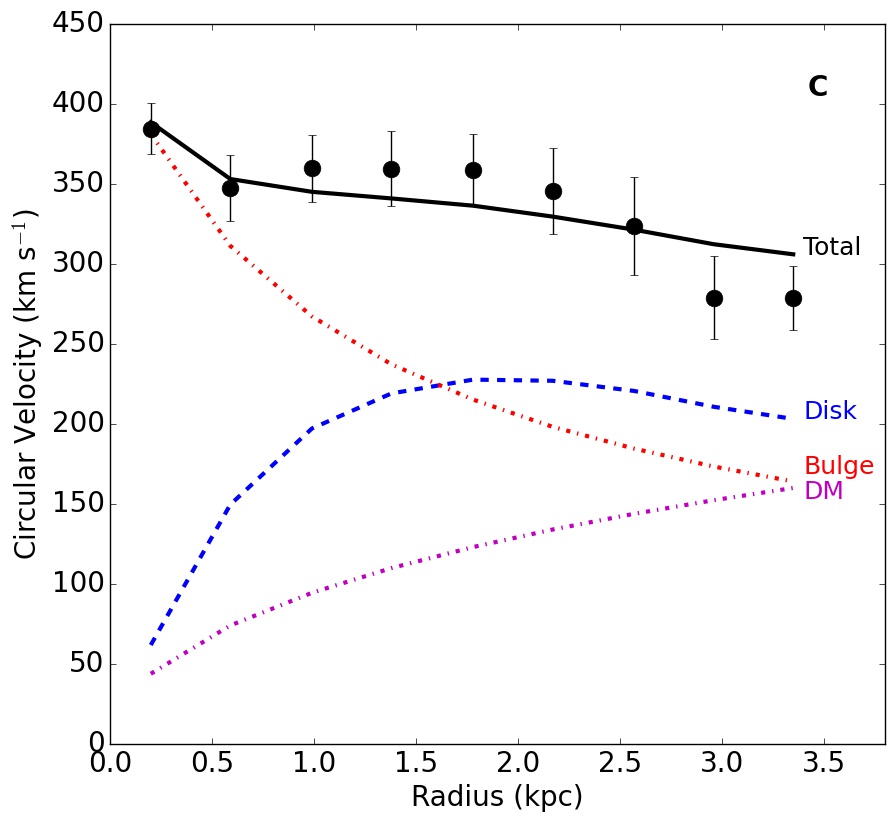}
\includegraphics[width=7.8cm]{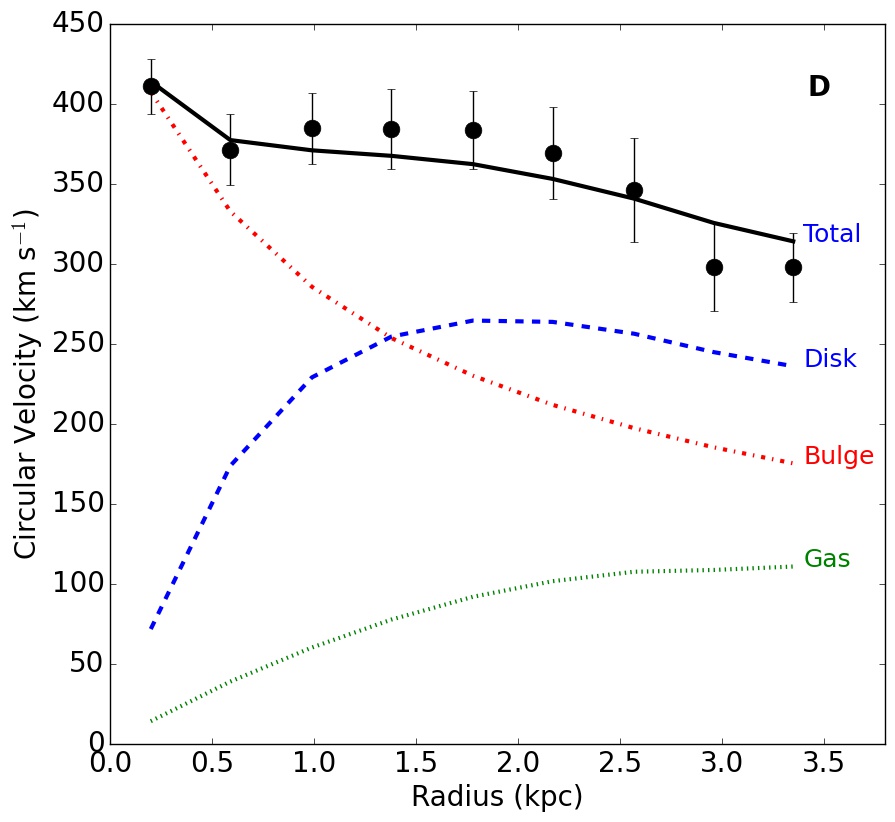}
\caption{\textbf{Alternative Mass Models:} the observed rotation curve (black dots) is compared with mass models fitted by assuming no bulge (\textbf{A}), no stellar disk (\textbf{B}), no gas (\textbf{C}), and no dark matter halo (\textbf{D}). The best-fitting model (black, solid line) in each case is the sum of gravitational contributions from (when present): stellar bulge (dot-dashed, red line), stellar disk (dashed, blue line), cold gas disk (green, dotted line), and DM halo (dash-dotted, magenta line).}
\end{figure*}

The model without the bulge (Fig. S5A) is unsatisfactory because the inner rotation velocities cannot be reproduced. The MCMC fitting tries to compensate for the lack of the bulge component by increasing the gravitational contribution of the stellar disk, as well as decreasing the total rotation velocities by pushing the inclination angle up to 23$^{\circ}$. This overshoots the rotation curve at large radii. A massive central component, which is not traced by the dust emission, is therefore necessary to explain the observed kinematics of ALESS\,073.1.

The model without stellar disk (Fig. S5B) provides a better fit but does not capture the observed shape of the rotation curve. This model assumes that the star-formation activity, traced by the dust emission, has just started in the disk and no substantial stellar component has been built yet. We regard this as unlikely because the estimated star-formation rate of 1000 M$_\odot$ yr$^{-1}$ can build a stellar disk of 10$^{10}$ M$_\odot$ in 10$^{7}$ years. This model gives $M_{\rm bul}/M_{\rm baryon} = 0.55$ which is higher than $M_{\rm bul}/M_{\rm baryon} = 0.44$ from our fiducial model because the relative contribution of the bulge stays almost the same as it is fixed by the inner rotation curve, while the relative contributions of gas and DM increase to explain the outer rotation velocities. $M_{\rm bul}/M_{\rm baryon}$ remains $>0.4$ confirming that a prominent bulge is in place at $z\simeq5$.

The model without gas disk (Fig. S5C) is almost identical to the fiducial mass model because the subdominant gas contribution is compensated by a slightly heavier stellar disk. This gives $M_{\rm bul}/M_{\rm baryon}=0.45$ which is equal to our fiducial value within the uncertainties (Table S2). We regard this model as unphysical because \cii\ and CO($J=2\rightarrow1$) emission lines are detected in ALESS\,073.1.

The model without DM halo (Fig. S5D) is similar to our fiducial model but has a lower value of $M_{\rm bul}/M_{\rm baryon}=0.37$. This indicates that DM within 3.5 kpc is not strictly necessary. This model is equivalent to a rotation curve in the context of Modified Newtonian Dynamics (MOND), which replaces DM with a departure from the classical laws of Newtonian dynamics at accelerations smaller than $a_0\simeq10^{-10}$ m s$^{-2}$ \cite{Milgrom1983a}. Assuming that the acceleration scale $a_0$ does not vary with $z$ and that the redshift-distance relationship in a MOND cosmology is similar to that in $\Lambda$CDM \cite{Skordis2020}, the observed accelerations in the outer regions of ALESS\,073.1 are about 7 times higher than $a_0$, so the galaxy is expected to behave as a classic Newtonian system with no mass discrepancy.

% For your review copy (i.e., the file you initially send in for
% evaluation), you can use the {figure} environment and the
% \includegraphics command to stream your figures into the text, placing
% all figures at the end.  For the final, revised manuscript for
% acceptance and production, however, PostScript or other graphics
% should not be streamed into your compliled file.  Instead, set
% captions as simple paragraphs (with a \noindent tag), setting them
% off from the rest of the text with a \clearpage as shown  below, and
% submit figures as separate files according to the Art Department's
% instructions.

\section*{Captions for Data}

\textbf{Data S1:} observed \cii\ data cube after cleaning and continuum subtraction. File in Flexible Image Transport System (FITS) format with a single layer.\\
\textbf{Data S2:} observed \cii\ intensity map (moment zero). File in FITS format with a single layer\\
\textbf{Data S3:} observed \cii\ velocity map (moment one). File in FITS format with a single layer.\\
\textbf{Data S4:} model \cii\ data cube from \textsc{$^{\rm 3D}$Barolo}. File in FITS format with a single layer.\\
\textbf{Data S5:} model \cii\ intensity map from \textsc{$^{\rm 3D}$Barolo}. File in FITS format with a single layer.\\
\textbf{Data S6:} model \cii\ velocity map from \textsc{$^{\rm 3D}$Barolo}. File in FITS format with a single layer.\\
\textbf{Data S7:} continuum maps from \textsc{Galfit}. File in FITS format with four layers: Ext[1] with the observed continuum map, Ext[2] with the model continuum map, Ext[3] with the residual continuum map. The primary layer (Ext[0]) is empty.\\
\textbf{Data S8:} best-fitting mass model. File in machine-readable table (mrt) format.

\bibliography{ALESS73}
\bibliographystyle{Science}

\end{document}